\documentclass[final,5p,times,twocolumn]{elsarticle} 

\usepackage{graphicx}
\graphicspath{ {pics/}}

\usepackage{threeparttable}

\usepackage{comment}
\usepackage{xcolor}

\usepackage[caption=false]{subfig}

\usepackage{lineno, hyperref}

\usepackage{amssymb}
\usepackage{amsmath}
\usepackage{amsthm}
\usepackage{amsfonts}
\usepackage{amssymb}

\usepackage{braket}

\usepackage{hyperref}
\usepackage{array}
\begin{document}

\begin{frontmatter}

\author[umic]{S. Marin\corref{cor1}}
\ead{stmarin@umich.edu}

\author[anl]{I. A. Tolstukhin\corref{cor1}}
\ead{itolstukhin@anl.gov}

\author[umic]{N. P. Giha}
\author[anl]{M. B. Oberling}
\author[anl]{R. A. Knaack}
\author[anl]{B. P. Kay}
\author[lanl]{D. L. Duke}
\author[lanl, mines]{K. B. Montoya}
\author[lanl]{D. Connolly}
\author[osu]{W. Loveland}
\author[osu]{A. Chemey}
\author[umic,umic2]{S. A. Pozzi}
\author[anl]{F. Tovesson}

\address[umic]{Department of Nuclear Engineering and Radiological Sciences, University of Michigan, Ann Arbor, MI 48109, USA}
\address[anl]{Physics Division, Argonne National Laboratory, Lemont, IL 60439, USA}
\address[osu]{Chemistry Department, Oregon State University, Corvallis OR 97331, USA}
\address[lanl]{Los Alamos National Laboratory, Los Alamos, NM 87545, USA}
\address[mines]{Department of Physics, Colorado School of Mines, Golden, CO 80401, United States}
\address[umic2]{Department of Physics, University of Michigan, Ann Arbor, MI 48109, USA}

\title{Instrumentation for correlated prompt $n$-$\gamma$  emission studies in coincidence with fission fragments}
\date{\today}


\begin{abstract}

Recent theoretical and experimental results have brought renewed interest and focus on the topic of fission fragment angular momentum. Measurements of neutrons and $\gamma$ rays in coincidence with fission fragments remain the most valuable tool in the exploration of fission physics. To achieve these scientific goals, we have developed a system that combines a state-of-the-art fission fragment detector and $n$-$\gamma$ radiation detectors. A new twin Frisch-gridded ionization chamber has been designed and constructed for use with a spontaneous fission source and an array of forty \textit{trans}-stilbene organic scintillators (FS-3) at Argonne National Laboratory. The new ionization chamber design we present in this work aims at minimizing particle attenuation in the chamber walls, and provides a compact apparatus that can be fit inside existing experimental systems. The ionization chamber is capable of measuring fission fragment masses and kinetic energies, whereas the FS-3 provides neutron and gamma-ray multiplicities and spectra. The details of both detector assembly are presented along with the first experimental results of this setup. Planned event-by-event analysis and future experiments are briefly discussed. 
\end{abstract}

\end{frontmatter}

\section{Introduction}
\label{sec:intro}

Recent papers by Wang \textit{et al.}~\cite{wang:2016}, Wilson \textit{et al.}~\cite{wilson:2021}, and Travar \textit{et al.}~\cite{travar:2021} have challenged several long-standing assumptions~\cite{wilhelmy:1972, glassel:1989} regarding the fragment angular momenta, including their mass dependence and the angular momentum correlations between the two fragments. Almost simultaneously with these experimental advances,  theoretical predictions and explanations have flourished in the literature. These have focused on the production mechanisms of angular momentum~\cite{randrup:2021, randrup:2022, bulgac:2021, bulgac:2022} and the sharing of excitation energy~\cite{schmidt:2011, bulgac:2020}. Ambitious microscopic calculations of fission observables have gathered interest in recent years~\cite{schunck:2022}, and their development will require both experimental results as well as their phenomenological interpretation.

In order to provide a complete picture of the fission process, measurements of neutrons and $\gamma$ rays emitted by the fragments immediately following fission are required. These radiated particles are the primary information carriers for the short-lived initial fragment states. In addition to the neutron and $\gamma$-ray multiplicities, it is also important to understand particle energies and angular distributions. Our recent papers~\cite{marin:2020, marin:2021, marin:2022} have shown that organic scintillator arrays provide insight into the fission process, revealing several features of the fragments' angular momenta, including their magnitudes, directions, and correlations between them. Previous event-by-event experiments performed were however limited in their lack of fragment mass and kinetic energy measurements. The purpose of this paper is to present the development of a new detection system designed for $n$-$\gamma$ event-by-event correlations studies in fission, which will reveal details about the relationship between fragment properties and their angular momentum. 

Frisch-gridded ionization chambers have been broadly used in nuclear physics applications, among them the measurement of fission fragments~\cite{budtz:1987}. Arrays of organic scintillators have been used to simultaneously measure neutrons and $\gamma$ rays emitted during the fission fragment de-excitation~\cite{marin:2022, gook:2014}. Combining the two systems allows us to correlate the emission of particles with fragment properties such as masses, kinetic energies, and ultimately, excitation energies. 


This work reports on the development and characterization of a flange-less Frisch-gridded ionization chamber combined with the FS-3 array of forty \textit{trans}-stilbene organic scintillators. We provide technical descriptions of the two instruments individually as well as the methodology and performance of the combined system. 

\section{Instruments}

\subsection{Twin Frisch-gridded ionization chamber (TFGIC)}

TFGICs have become popular in fission studies, and there is an extensive literature describing their function and modes of operation~\cite{budtz:1987, gook:2014, gaudefroy:2017, mosby:2014}. The TFGIC design used in this experiment was inspired by the design by Dana Duke~\cite{duke:thesis}, with a few significant modifications to the vacuum chamber surrounding the detector and the readout boards to reduce the attenuation of neutrons and $\gamma$ rays.

\begin{figure}[!htb]
\centering
\includegraphics[width=\linewidth]{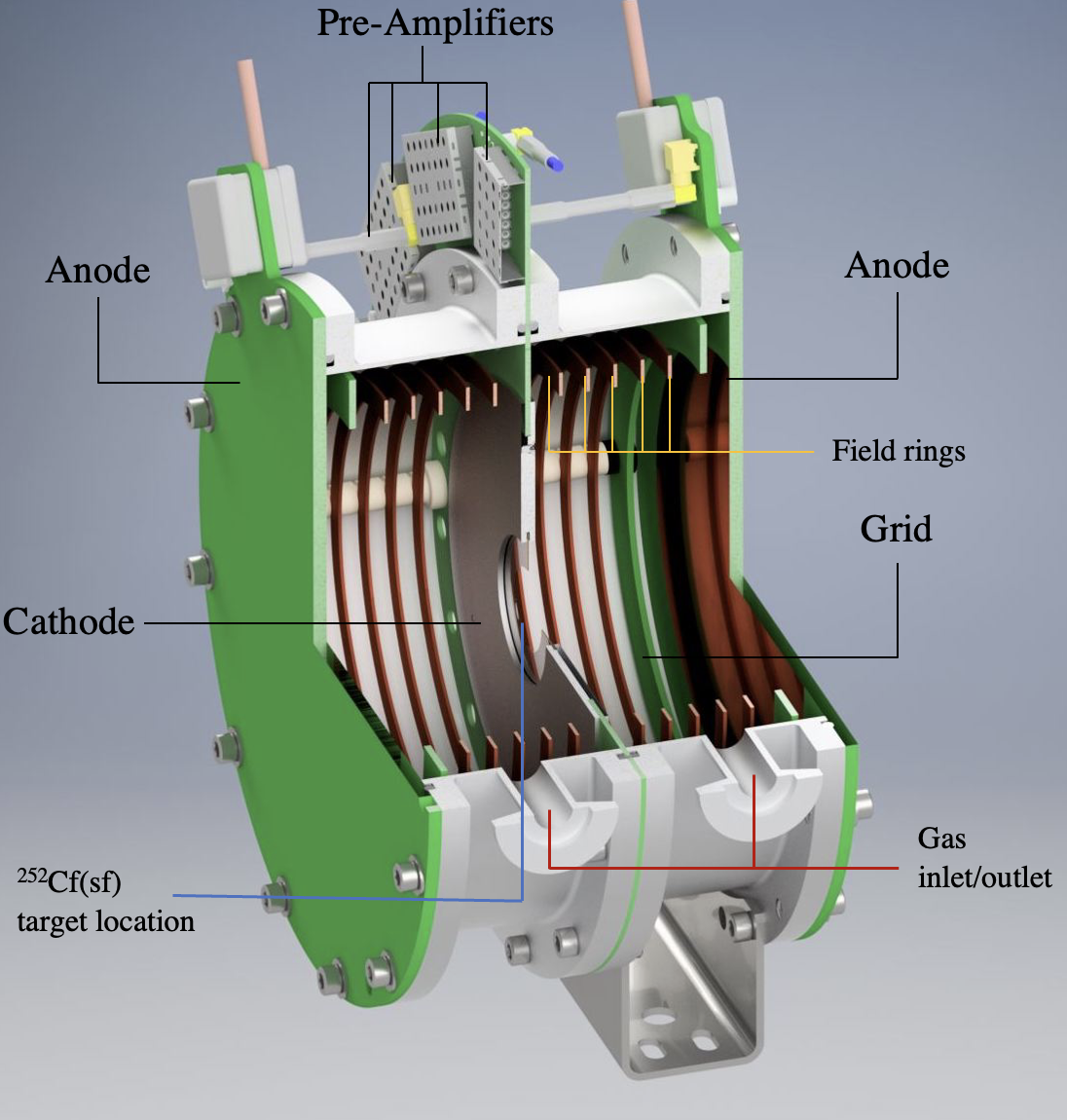}
\caption[Ionization Chamber Diagram]{CAD diagram of the TFGIC.}
\label{fig:TFGIC}
\end{figure}

A CAD drawing of the TFGIC is shown in Fig.~\ref{fig:TFGIC}. The fragment detector is composed of two identical volumes enclosed between the central cathode plate and the two anode plates. The inner diameter of the chamber is $140~$mm and the distance between cathode and anode boards is $47~$mm. The anode and its associated circuitry are combined in a single printed circuit board (PCB). Similarly, the cathode and the associated electronic circuitry, as well as the preamplifiers for all the TFGIC signals, are contained on a single PCB. To minimize neutron and $\gamma$-ray attenuation, the anodes' circuit boards are used to enclose the chamber volume, thus avoiding the use of metal flanges. 





Frisch grids are used in ionization chambers to eliminate the geometric dependence of the anode signal, and to improve the timing resolution of the detector~\cite{knoll:2010}. In this experiment, awe have the conventional approach of using the ratio of grid to anode signals to extract the polar angle of emission of fragments with respect to the chamber axis. The grids are made of $20 \ \mu\text{m}$ gold-plated tungsten wires spaced $1 \text{~mm}$ apart and soldered to PCB disks. These grids are placed between the cathode and each anode at a distance of $7\text{~mm}$ from the anode.

The anodes are 4 layer PCBs constructed from FR4 laminate substrate with an active area diameter of 108 mm. The anodes are $3.175$ mm thick to support operation with a differential working pressure of up to $0.3$~atm.  The anode uses $1$~oz. copper, $34.8~\mu \text{m}$ thick, on the outer conductive surface that is finished with an electroless nickel immersion gold (ENIG) process. A guard ring energized to the same potential as the anode surface encircles the anode to improve field uniformity at the edge of the anode's active area. Each anode provides a shielded HV hookup that includes an HV filter, signal decoupling capacitor, and the necessary biasing resistors for the anode and its guard ring.  The anode signal is connected via a short $\sim 5$ cm cable to the preamplifier located on the cathode. Collecting all the preamplifiers on the cathode board keeps them in close proximity, while simplifying the electronics and cabling, and also allows the anode design to be modified and swapped out more freely than if they were directly on PCB.

The ionization chamber volume is filled with P-10 gas, Ar(90\%)+CH$_4$(10\%) at $950~$torr with continuous flow of $\sim 100~$cc/min. Each section of the twin chamber has a gas port, which are respectively used as inlet and outlet. To facilitate the gas circulation between the two sections, 8 holes of $6.5$~mm diameter are located on the cathode board on the opposite side of the gas ports. The gas pressure is monitored, and variations on the order of $10$ torr have been observed, but the electrodes' signals were not significantly influenced by these small variations.

The TFGIC detector volume is electrified by holding the cathode at a potential of $-1500$ V, the two grids grounded at $0$ V, and the anode plates at $+1000$ V. These voltages were provided by CAEN N1470 power supplies. The produced electric field is rectified and made uniform across the chamber through the use of copper \textit{field rings}, five in each chamber section. The field rings are held by three PEEK columns that are mounted directly on the cathode board. 

\begin{figure}[ht]
\begin{center}
\includegraphics[scale=0.35]{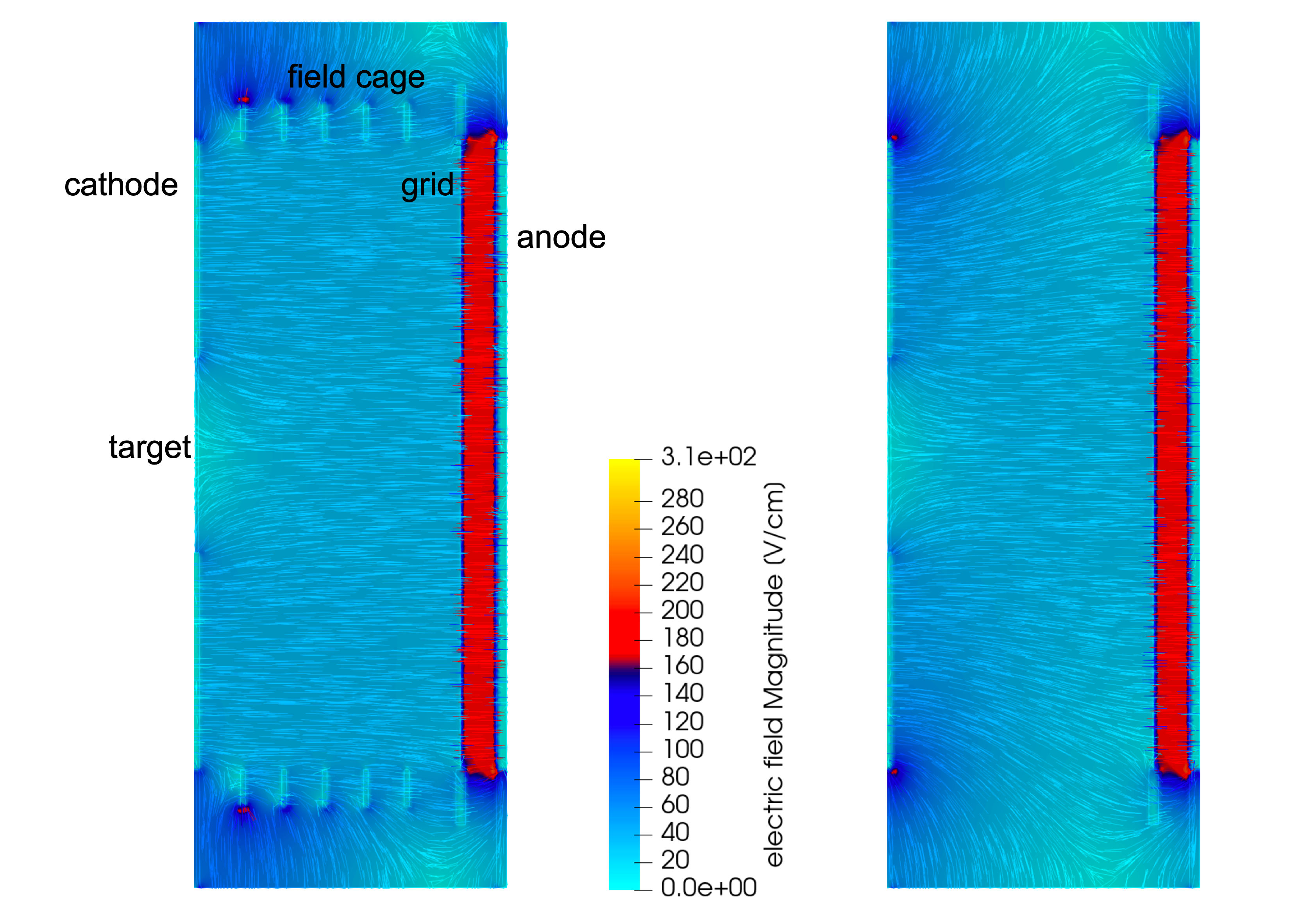}
\caption{Calculated magnitude of the electric field with (left) and without (right) field cage rings.}
\label{fig:tfgic_field}
\end{center}
\end{figure}

The Gmsh finite-element mesh generating software~\cite{GMSH:2009} was used to develop the geometry and to perform a mesh generation throughout the volume of the ionization chamber. Due to the symmetry, only half of the chamber is used for modeling the detector. 
The uniformity the electric field inside the chamber was investigated using the Elmer finite-element software~\cite{elmer_soft} and a previously generated mesh. The electrostatic problem is defined by assigning the dielectric properties of the materials in each sub-volume. The calculated magnitude of the electric field and orientation inside the detector are shown for two cases with (Fig.~\ref{fig:tfgic_field}, left) and without (Fig.~\ref{fig:tfgic_field}, right) field rings.

The five signals generated by the TFGIC: one cathode, two grids and two anodes, are passed through Cremat CR-110 preamplifiers mounted directly on the cathode board, outside of the aluminum walls of the chamber. The preamplified signals have a short rise time, around $ 200-250$ ns, and a long decay time of $150~\mu \text{s}$. The grid signals are digitized using a CAEN V1740, 64-channel digitizer, with 12-bit resolution over a $2$ V dynamic range and a $62.5$-MHz sampling rate. The signals from the anodes and the cathode are cloned using a CAEN N454 fan-in/fan-out module, with one of the copies of each channel being digitized in the V1740. Clones of both anode signals are provided in channels 0 and 1 of the three V1730 digitizers, for coincidence purposes. A clone of the cathode signal is provided to one of the V1730 digitizers \textemdash the FS-3 signal digitizers \textemdash, where a digital CFD algorithm determines its  timing. We have determined a time resolution of $\approx 5-6 $ ns FWHM from the broadening of the $\gamma$-ray coincident timing spectrum. Another copy of the cathode signal is provided to an oscilloscope for use as a diagnostic. 

 We have designed an aluminum bracket and holder system, that allows the chamber to be vertically repositioned and rotated. The chamber was aligned with the source being at the geometric center of the FS-3 array, and with the axis of the chamber\textemdash the line of shortest distance between cathode and anode\textemdash pointed in the direction of one of the trans-stilbene detectors. 

A spontaneous fission source was prepared by molecular plating of $9$~kBq of $^{252}$Cf on a $\sim100 \ \mu\text{g/cm}^2$ carbon foil at Oregon State University. The diameter of the deposit on the backing is $10$~mm, and it was determined that the source was geometrically offset by about $2$~mm with respect to the center of the carbon foil. This offset was deemed negligible, since this distance is much shorter than both typical fragment ranges and the dimension of the detector active volume.

Data were collected from the detectors and TFGIC only when signals from both anodes were observed in coincidence, \textit{i.e.}, the trigger condition. This coincidence AND logic significantly lowers the background, and virtually eliminates the $\alpha$-particle background, as can be determined by pulse-height spectroscopy and comparison of the chamber throughput to the nominal source activity.

\subsection{FS-3 Array}

\begin{figure}[!htb]
\centering
\includegraphics[width=0.8 \linewidth]{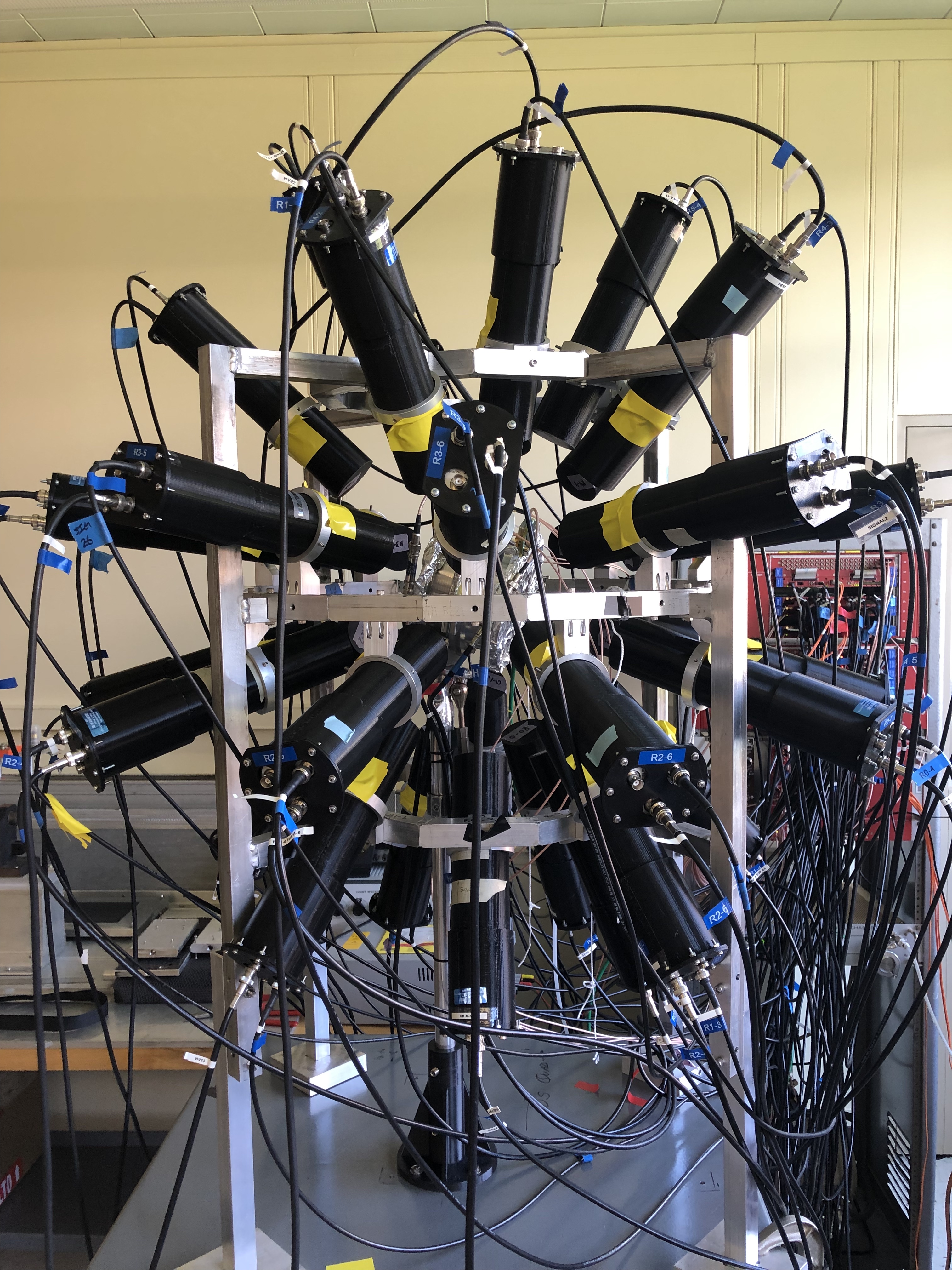}
\caption[FS-3 detector array]{The FS-3 detection array at Argonne National Laboratory. The TFGIC is shown at the center the detector array. Also visible (background, right) is the electronic readout and high-voltage supply. }
\label{fig:FS3}
\end{figure}

The FS-3 detector array, shown in Fig.~\ref{fig:FS3}, consists of forty organic scintillator detectors arranged in spherical configuration.  Each detector consists of a 5.08 cm by 5.08 cm  right circular cylinder \textit{trans}-stilbene crystal manufactured by Inrad-Optics to our specifications~\cite{shin:2019}. Each crystal is optically coupled to a ElectronTube 9214B photo-multiplier tube (PMT), and are individually wrapped in insulating tape and teflon to reduce optical noise, and mu-metal to reduce the effects of external magnetic fields. Finally, the assembly is placed inside a 3D printed case, which further reduces optical noise and allows it to be easily handled. 

The detectors are arranged in a spherical configuration, with detector holders placed around three concentric rings. The rings are held in place by aluminum columns, which have been designed to be modular and thus allow for changes in the height of the detector array. The support structure allows each detector to be placed at a variable distance from the center of the source, independently from one another, from a minimum of 14 cm up to 27 cm. In the present experiment, an intermediate distance of 22.5 cm was used. A detailed model of the detectors, the aluminum structure, and the surrounding room has been generated for MCNPX-PoliMi~\cite{pozzi:2003, pozzi:2012}. The FS-3 detectors are individually powered using seven CAEN V6533 negative polarity power supplies. The power supplies are connected via USB to the DAQ, and are operated using the CAEN GECO2020 control software. The high voltage (HV) on each PMT is adjusted to calibrate all detectors on the Compton edge of a $^{137}$Cs source. The calibration was repeated daily, but only minor corrections, on the order of $2 \%$ in the individual detector calibrations, were observed after the detectors reached thermal equilibrium. The signal of each detector is individually digitized using CAEN V1730 digitizers, with 500 MHz digitization rate, and $2$ V dynamic range. Each digitizer reads out 16 channels, and three V1730 digitizers were used in this experiment. The three digitizer clocks are synchronized with one another and with the clock of the V1740 digitizer collecting the TFGIC data.

\subsection{DAQ and signal processing}

\begin{figure}[ht]
\begin{center}
\includegraphics[scale=0.25]{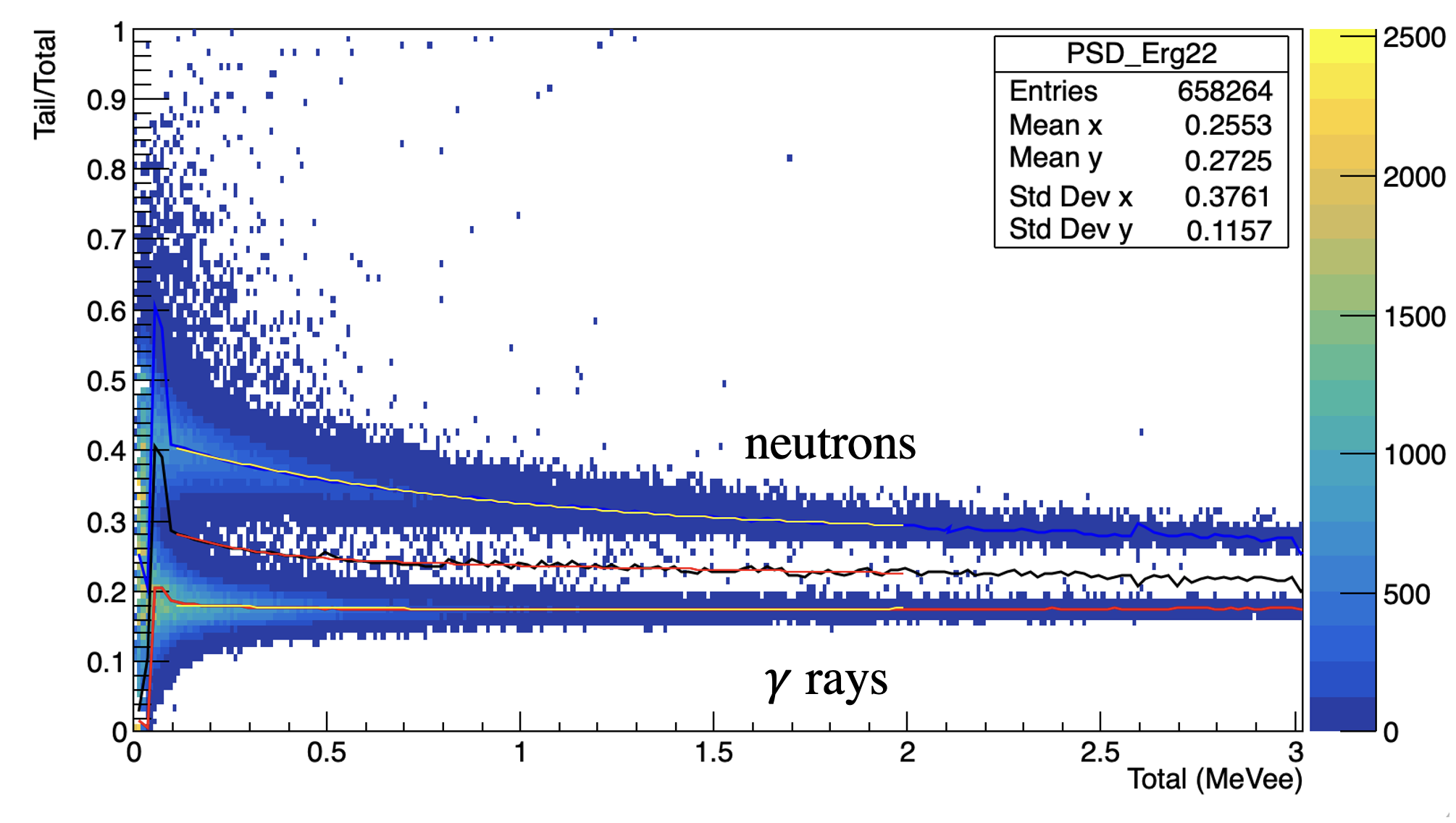}
\caption{Neutron and $\gamma$-ray PSD from one of the detectors in FS-3. Neutrons and gamma rays separate in two bands when we take the ratio of the integral of the tail of the scintillation pulse to the total integral. Energy dependent discrimination and Gaussian fits are shown. }
\label{fig:psdDet}
\end{center}
\end{figure}

The FS-3 detector signals are analyzed on the digitizer boards using charge integration. The time integral of the voltage signals is proportional to the light output generated by the interaction. The total light output is calibrated using the Compton edge of $^{137}$Cs, which provides a conversion of the light output to energy deposited. We measure light output in units of eV electron-equivalent, or eVee. The protons that are scattered in neutron interactions generate significantly less scintillation light and a greater portion of this light is produced as delayed scintillation~\cite{birks:1964}. By comparing the amount of scintillation light produced a few ns after the interaction, more characteristic of delayed fluorescence, to the total light output, we can distinguish interactions caused by $\gamma$ rays and neutrons on an event-by-event basis. This procedure is known as pulse shape discrimination (PSD). A PSD plot is shown in Fig.~\ref{fig:psdDet}. We have written an algorithm that computes the optimal discrimination as a function of the total light output, and we optimized the region of integration for the tail of the pulse, where delayed fluorescence is expected to be stronger. 

A threshold of 50 keVee is applied to the scintillator signals. This threshold results in a $\gamma$-ray incident-energy threshold of $\gamma$-ray of $0.15$ MeV, but a larger threshold of approximately $0.5$ MeV. The larger neutron threshold results in a bias in this experiment to observe neutrons of higher energies.   

\begin{figure}[ht]
\begin{center}
\includegraphics[scale=0.4]{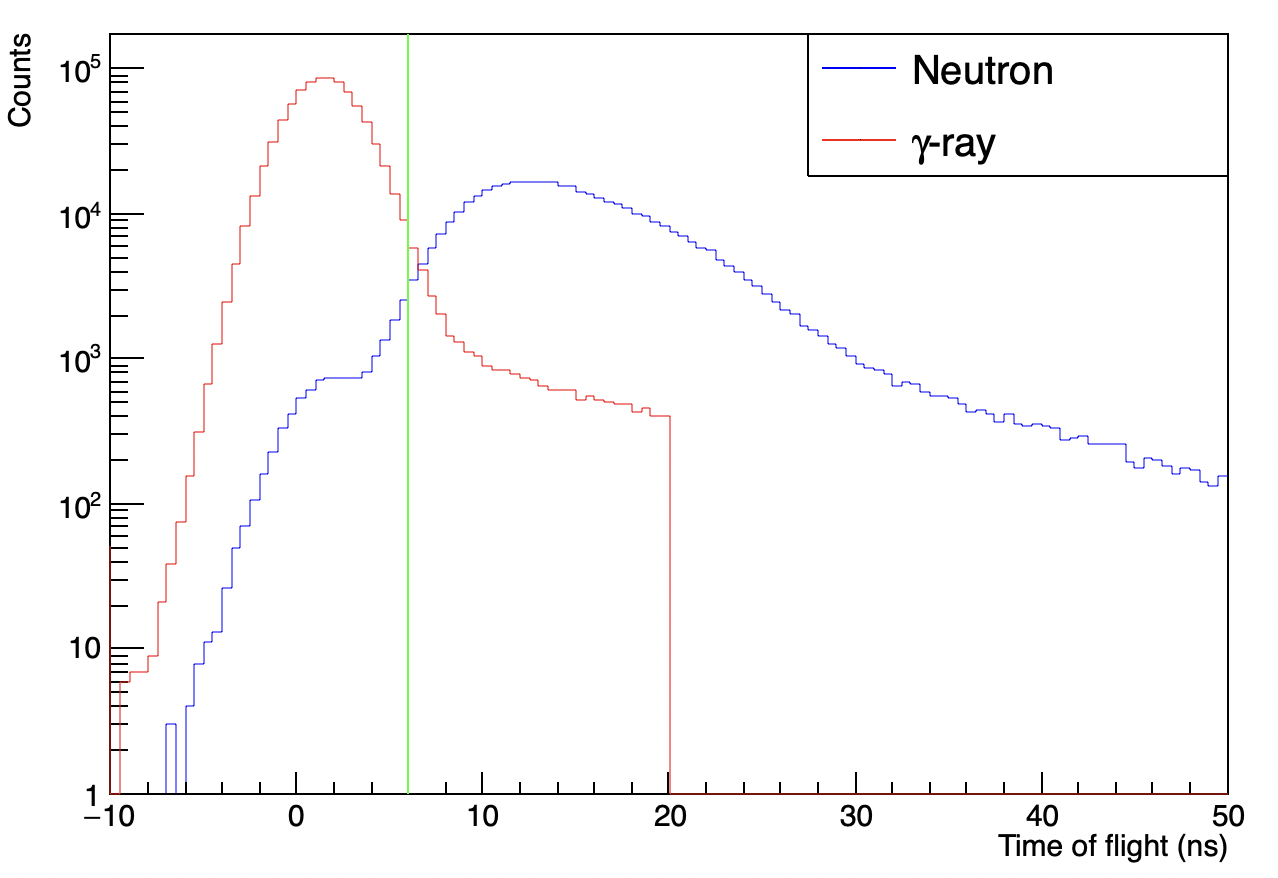}
\caption{Time-of-flight distribution from the TFGIC to a detector in FS-3. Neutrons and $\gamma$ rays are discriminated based on both PSD and ToF. Limited neutron ToF spectroscopy can be applied to the measured neutrons. The vertical green line, at $6$ ns shows the timing separation between $\gamma$ rays and neutrons. }
\label{fig:tofDet}
\end{center}
\end{figure}

The system's time resolution is good enough to allow the use of ToF for particle classification and, to a limited degree, for neutron spectroscopy. The ToF distribution of the measured particles, with respect to the measured cathode time, is shown in Fig.~\ref{fig:tofDet}. The simultaneous use of both PSD and ToF for particle classification results in a negligible misclassification rate. In fact, the neutrons that arrive the earliest,  simultaneosuly with $\gamma$ rays, are also the most energetic neutrons, the easiest to discriminate using PSD. 

\section{Analysis}

We analyse the TFGIC signals using the $2E$ method, which determines the masses of the fission fragment based on the measurement of the two fragment kinetic energies and the law of conservation of angular momentum. The $2E$ procedure has been presented in several past publications; in the following we will briefly summarize the main analysis steps, and focus on the improvements we have performed with this procedure. We refer the reader to Refs.~\cite{gook:2014, duke:thesis, moore:2021} for detailed descriptions of this technique. 

The analysis begins with the reconstruction of the fission fragment kinetic energy from the anode signal. The stray signal induced on the anode by charges drifting between cathode and grid, an effect known as \textit{grid inefficiency}~\cite{gook:2014}, is corrected for by comparing the event-by-event signals of anode and grid from the same side of the chamber. Secondly, the energy lost by the fragments in the carbon backing and within the source deposit itself is estimated. This estimate is based on the reconstructed fragment angle, as determined from the ratio of grid signals to anode signals~\cite{budtz:1987}. Finally, the fragment masses are recursively determined by determining, at each step in the recursion, to the neutron multiplicity and the pulse height defect (PHD).

We have introduced, in our analysis, an additional step of recursion by repeating all the above steps, but differentiated with respect to the fragment masses determined in the previous recursion step. This recursion aids in the analysis of angles, where differences in fragment charges can cause differences in the range of fragments in the P-10 mixture. The mass-differentiated analysis also improves the correction for the fragment attenuation, which improves both energy and mass resolution. 

The quantities of interest we want to extract from the ionization chamber are the yield observables: the fragment masses, $A$, and the total kinetic energy release, TKE. However, while not directly reported, the angle of emission of the fragments with respect to the TFGIC cylindrical symmetry axis is an important auxiliary variable in the analysis of fragment features. Specifically, the fragment angle determines the corrections that need to be applied to correct for the attenuation of the fragments in the target backing. 

\begin{figure}[!htb]
\centering
\includegraphics[width=0.8 \linewidth]{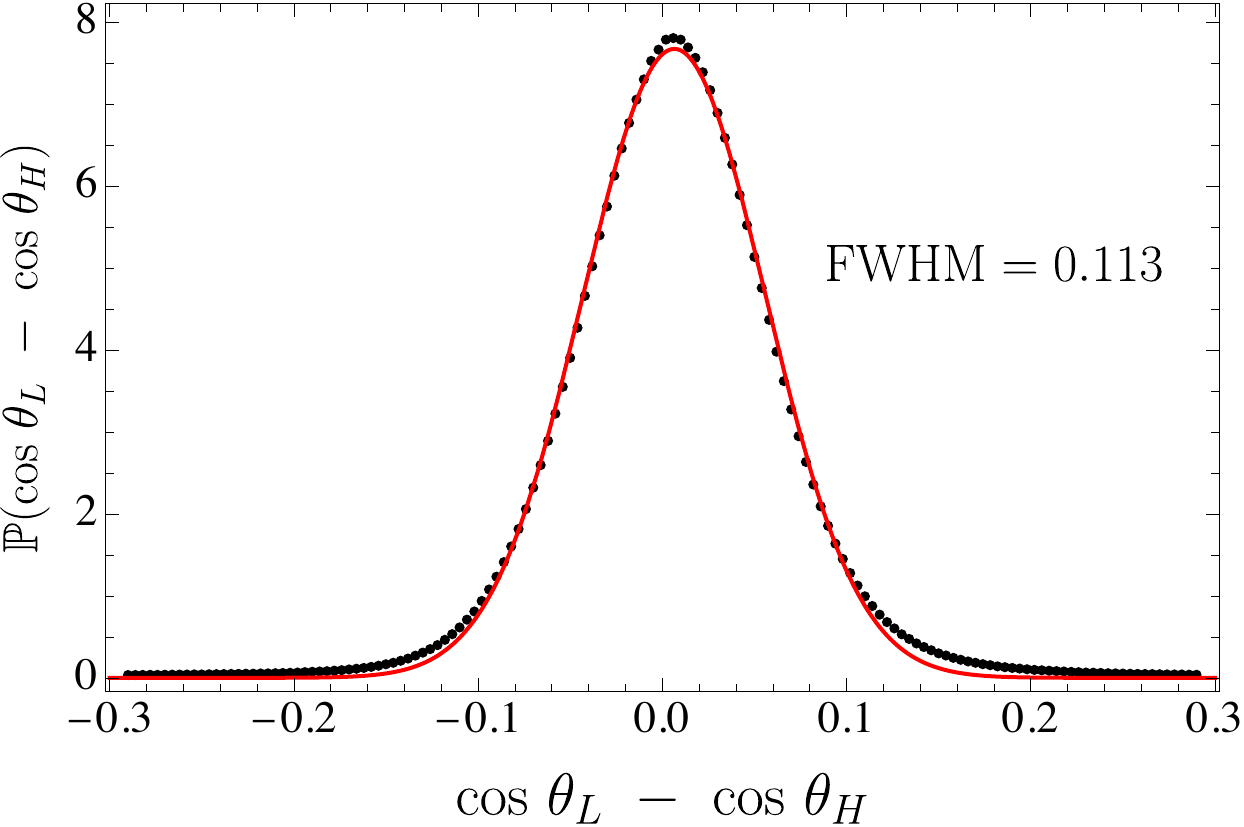}
\caption[Angular Resolution TFGIC]{Difference between the angles determined by either side of the chamber, which ideally would be identical. The width of this distribution is indicative of the angular resolution of the TFGIC. The red line is an illustrative Gaussian fit applied to the data }
\label{fig:AngResTFGIC}
\end{figure}

The fragment angle-of-emission can be determined by the ratio of the signal induced on the grid to the signal induced in the anode. Thus, the angle can be determined independently by each side of the fission chamber. Because fragments in spontaneous fission are emitted back-to-back, the variations between the two independent measurements can be used to assess the resolution of the TFGIC to the fragment direction. Fig.~\ref{fig:AngResTFGIC} shows the difference between the angle determined from the two sides of the chamber. An angular resolution of $0.11$ FWHM in cosine bins was determined, approximately 27 degrees. However, by combining the information from both sides of the chamber, the angular resolution can be reduced by half. 

The kinetic energy of fission fragments is determined primarily by the signal induced on the anodes. These signals are corrected for grid inefficiencies, the angle-dependent energy loss in the $^{252}$Cf sample, and its backing, and the pulse height defect in P-10. The kinetic energy is further corrected by reconstructing the energies prior to neutron emission using the mean value $\langle N | A, \text{TKE} \rangle$ determined by G\"o\"ok~\textit{et al.}\cite{gook:2014}. However, the mass $A$ is determined by comparing the fragment kinetic energies, and thus masses and kinetic energies are simultaneously determined in a recursive loop, as explained above. The recursive loop was interrupted when masses differed by less than $0.2$ \% between iterations. 

Due to the large attenuation of the fragments in the target and its backing, we find that data still contain angle dependence in the kinetic energy distributions, even after these effects are addressed with the method indicated in Ref.~\cite{duke:thesis}. To avoid these problems, we selected a narrow range of angle of emissions, $|\cos \theta| > 0.9$, where the angle is determined from the arithmetic average of the angles determined from each side independently. 

\begin{figure}[!htb]
\centering
\includegraphics[width=0.8 \linewidth]{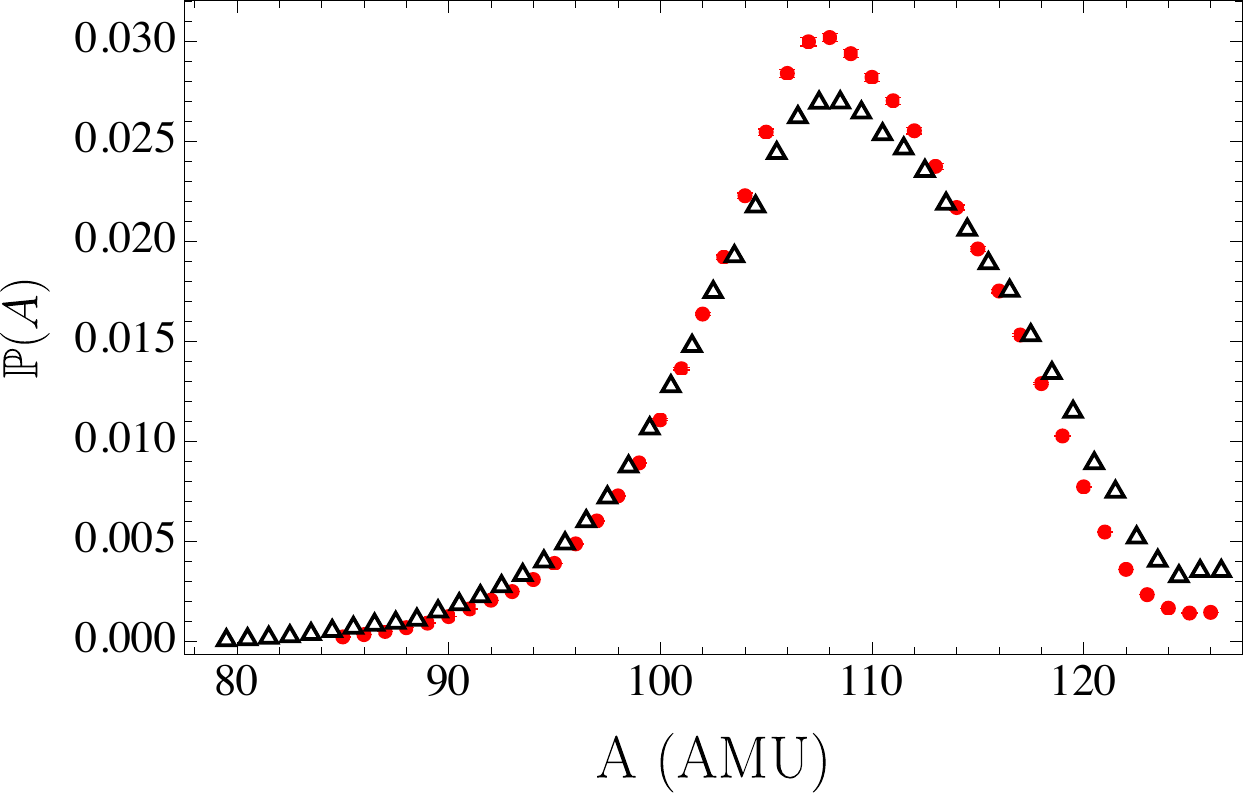}
\caption[Mass Resolution TFGIC]{Fragment mass yield determined by the TFGIC, black points (triangle), compared to the yield determined by G\"o\"ok~\textit{et al.}, shown as red points (circle). }
\label{fig:MassResTFGIC}
\end{figure}

The mass resolution of the TFGIC is shown in Fig.~\ref{fig:MassResTFGIC}, where it is compared to the data obtained by G\"o\"ok~\textit{et al.}\cite{gook:2014}. We note that because of symmetry, we only need to plot the yield as a function of the light fragment mass, as the same yield would be observed, by definition, for the complementary $A_0 - A$, where $A_0 = 252$ is the fissioning nucleus mass number. The agreement between the two experiments is quite good across the mass yield, with some deviations near the symmetric fission, $A \approx 120$. Generally, our distribution is slightly larger than the one inferred by G\"o\"ok~\textit{et al.}, indicating a worse mass resolution, approximately $4-5$ AMU FWHM. 

\begin{figure}[!htb]
\centering
\includegraphics[width=0.8 \linewidth]{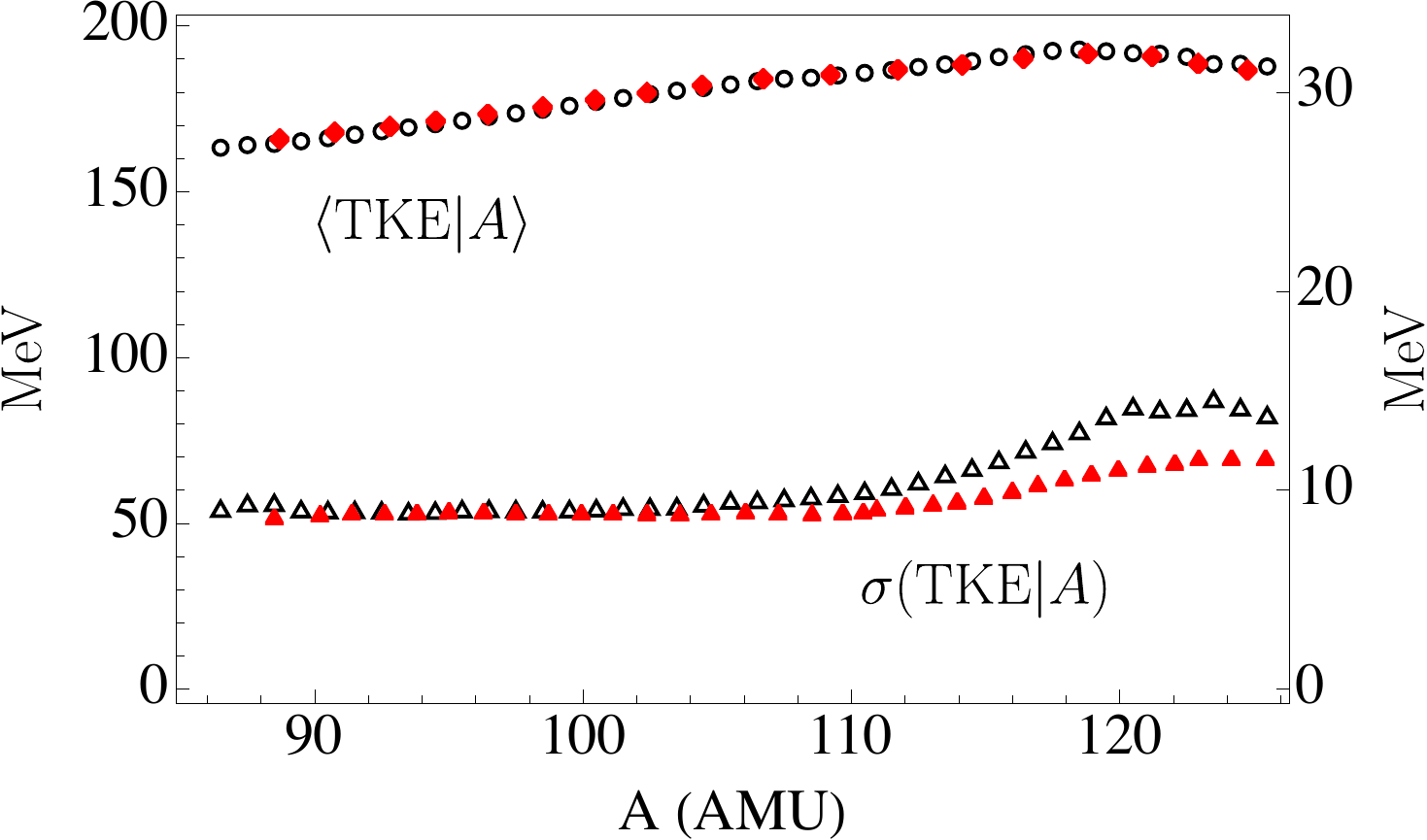}
\caption[Kinetic Energy Resolution TFGIC]{Average and standard deviation of the TKE distribution determined by the TFGIC, black points, compared to the yield determined by G\"o\"ok~\textit{et al.}, shown as red points.}
\label{fig:ErgResTFGIC}
\end{figure}

The fission TKE, conditioned on the light fragment mass is shown in Fig.~\ref{fig:ErgResTFGIC}. The figure shows both the mean and the standard deviation of the determined kinetic energy release, with the former indicative of the accuracy of the TFGIC, and the latter indicative of its kinetic-energy resolution. The determined mean $\langle \text{TKE} | A \rangle$ was found to be in good agreement with G\"o\"ok~\textit{et al.}, with slight deviations at $A \approx 120$ and $A < 95$. The width of the TKE distribution is comparable to the reference experiment throughout most of the mass yields, but it is larger near symmetric fission. These results indicate a kinetic energy resolution of approximately $3-4$ MeV FWHM. 

\section{Results}

\begin{figure}[!htb]
\centering
\includegraphics[width=0.8 \linewidth]{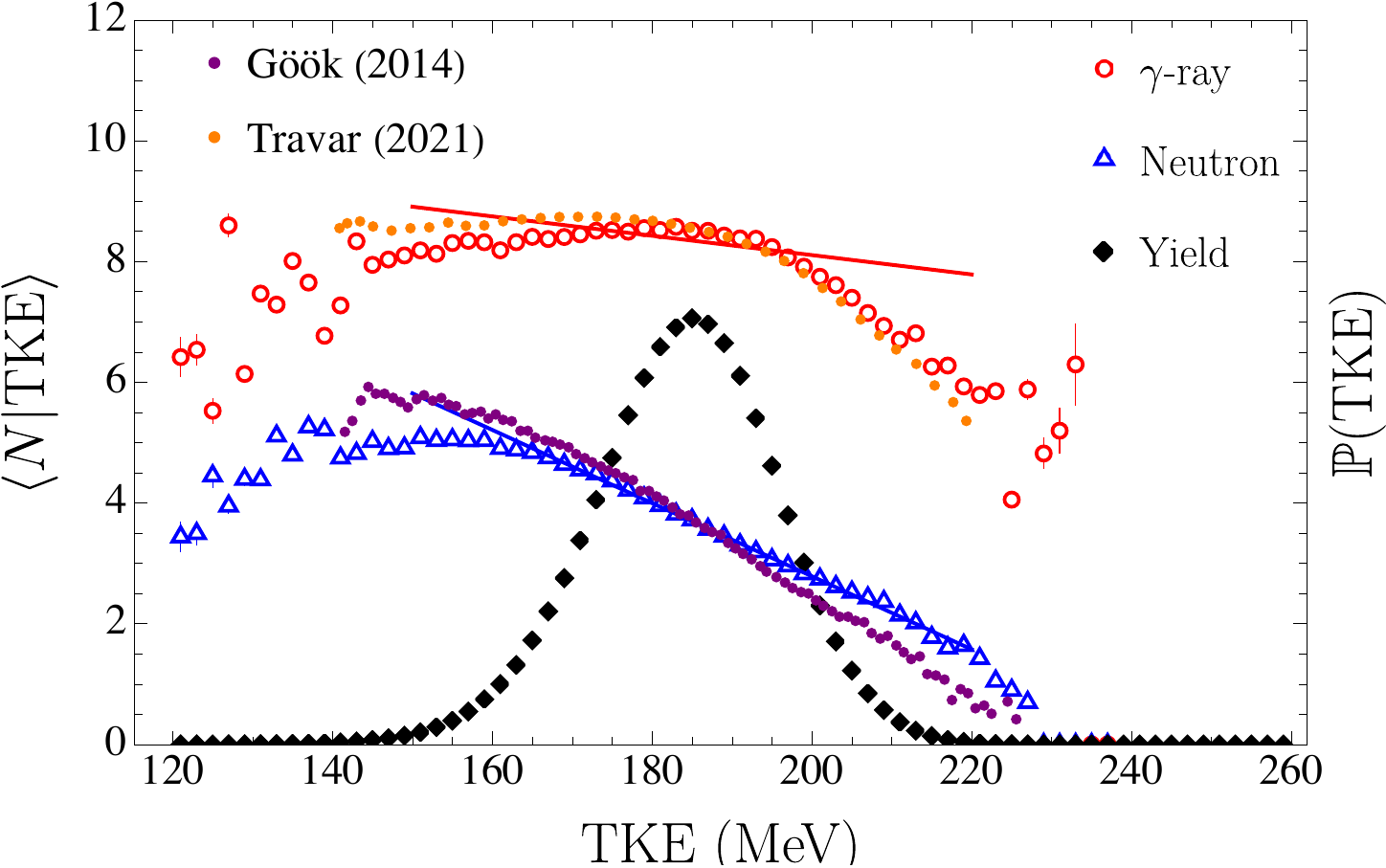}
\caption[]{Total neutron- and $\gamma$-ray-multiplicity dependence on fragment TKE compared to previous work by G\"o\"ok \textit{et al.}~\cite{gook:2014} and Travar \textit{et al.}~\cite{travar:2021}, respectively. The measured TKE yield is shown in black.}
\label{fig:tkeFragComp}
\end{figure}

\begin{figure}[!htb]
\centering
\includegraphics[width=0.8 \linewidth]{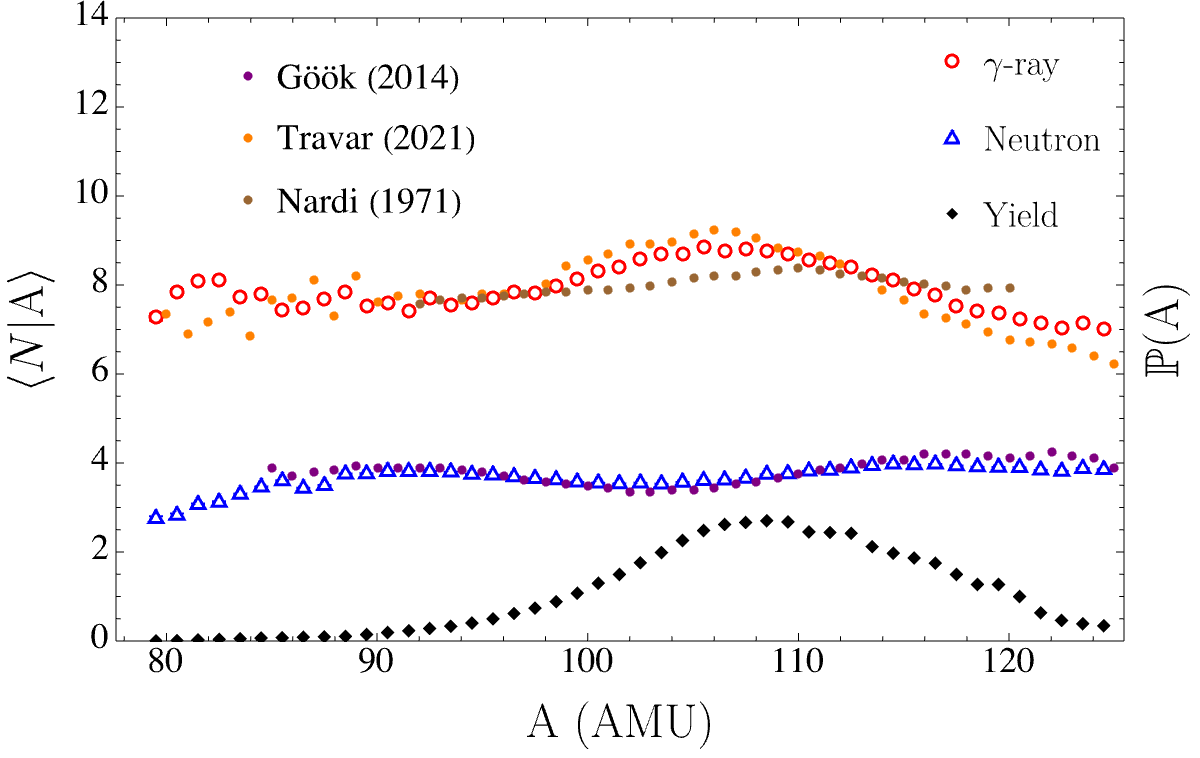}
\caption{Total neutron- and $\gamma$-ray-multiplicity dependence on the light fragment mass compared to previous work by G\"o\"ok \textit{et al.}~\cite{gook:2014} and Travar \textit{et al.}~\cite{travar:2021}, respectively. The measured mass yield is shown in black.  }
\label{fig:massFragComp}
\end{figure}

As an illustrative example of the results that this combined system can produce, we present here the conditional differentiation of neutrons and $\gamma$ rays with respect to total kinetic energy and fragment mass, in Figs.~\ref{fig:tkeFragComp} and \ref{fig:massFragComp} respectively. On the same figures as our experimental results, we compare the neutron emission results to G\"o\"ok \textit{et al.}~\cite{gook:2014} and Travar \textit{et al.}~\cite{travar:2021}, for neutrons and $\gamma$-ray results, respectively. Both of these previous results used a very similar setup, employing a TFGIC in coincidence with radiation detectors, but only one particle type was analyzed in each of those experiments. The results of this comparison show that due to the resolution achieved by our system so far, slightly larger than the resolutions achieved by G\"o\"ok \textit{et al.}, the features of the multiplicity distributions are slightly broadened, and the correlations of particle multiplicities with fragment masses and kinetic energy are slightly weakened.

When conditioned on TKE, we observe the same behavior already observed by Travar~\textit{et al.}, the $\gamma$-ray multiplicity increases with decreasing TKE until TKE $\approx 180$ MeV, below which the $\gamma$ ray multiplicity stops growing and levels off, and even starts to slightly decrease. 

While improvements can still be made to the fragment detection system, the results shown in Figs.~\ref{fig:tkeFragComp} and \ref{fig:massFragComp} illustrate that coincident measurements of neutrons, $\gamma$ rays, and fission fragments are possible with the system presented here. 

\section{Conclusion}

The experimental setup we presented in this paper will be used in the immediate future to analyse the event-by-event $n$-$\gamma$ emission correlations in fission in coincidence with measurements of fragment masses and excitation energy. These results will provide insight into the questions of fragment angular momenta and energy sharing. This system improves on the current available technologies by greatly increasing the number of available detectors, forty in our experiment, and by simultaneously gaining access to both neutrons and $\gamma$ rays in coincidence with fragments. Using this capability, we will explore the event-by-event emission competition of neutrons and $\gamma$ rays, and the effects that angular momentum has on this competition. 

Several alternative setups and improvements are planned for the assembly. We will soon perform a measurement with the TFGIC inside Gammasphere~\cite{LEE1990c641} at the Argonne National Laboratory ATLAS facility, an array of Compton-suppressed, high-purity germanium detectors, to obtain detailed spectroscopic and angular information correlating $\gamma$ rays with fragment properties.

We plan to segment one of the anodes of the TFGIC to obtain a measurement of the azimuthal angle of the fission axis, completely constraining its direction. This iteration of the TFGIC, coupled with both measurements within FS-3 and Gammasphere, will yield some of the most complete data on fission correlations to date. In future research, we also plan to use these instruments to investigate the event-by-event $n$-$\gamma$ correlations in neutron-induced fission, leveraging the recent investigation of fragment yields in these reactions~\cite{chemey:2020, pica:2020, yanez:2014}.

\section{Acknowledgements}

The authors would like to thank L. Yao and S. Scott for their work in the preparation of the experiment target. S.M. thanks M. S. Okar for contributions to the original design of the FS-3. This work was in part supported by the Office of Defense Nuclear Nonproliferation Research and Development (DNN R \& D), National Nuclear Security Administration, US Department of Energy. This work was funded in-part by the Consortium for Monitoring, Technology, and Verification under Department of Energy National Nuclear Security Administration award number DE-NA0003920.This material is based upon work supported by the U.S. Department of Energy, Office of Science, Office of Nuclear Physics, under Contract Number DE-AC02-06CH11357.

\bibliographystyle{elsarticle-num}
\bibliography{ref}

\end{document}